# Stable Asymmetric Magnetization Reversal in Epitaxial Co(001)/CoO(001) Bilayer

Maik Gaerner[1,‡], Judith Bünte[1,‡], Finn Peters[1], Inga Ennen[1], Hermann Tetzlaff[2], Johannes Fiedler[2], Tomasz Blachowicz[3], Luana Caron[1,4], Andreas Hütten[1], Andrea Ehrmann[2], Martin Wortmann[1,2,*]

[1] Bielefeld University, Faculty of Physics, Universitätsstraße 25, 33615 Bielefeld, Germany
[2] Bielefeld University of Applied Sciences and Arts, Faculty of Engineering and Mathematics, Interaktion 1, 33619 Bielefeld, Germany
[3] Silesian University of Technology, Institute of Physics—Center for Science and Education, 44-100 Gliwice, Poland
[4] Helmholtz-Zentrum Berlin für Materialien und Energie, 12489 Berlin, Germany
[‡] These authors contributed equally
[*] Correspondence: mwortmann@physik.uni-bielefeld.de

## Abstract

The exchange bias (EB) in ferromagnetic/antiferromagnetic (FM/AFM) bilayer systems causes a shift of the magnetic hysteresis curve after field cooling through the Néel temperature of the AFM. In some cases, this shift is accompanied by an asymmetry between ascending and descending branches. In the past, this asymmetric magnetization reversal has been studied in different bilayer systems, including polycrystalline Co/CoO thin films. Here we investigate the asymmetric magnetization reversal in an epitaxial fcc-Co(001)/CoO(001) thin film grown on MgO(001) by molecular beam epitaxy at varying temperatures up to room temperature after field cooling along the easy and the hard axes. Room temperature measurements of the longitudinal and transverse magneto-optic Kerr effect show different magnetization reversal processes via stable intermediate states for different angles between external magnetic field and magnetic easy axes. Once the sample is cooled below the blocking temperature, a pronounced asymmetric magnetization reversal can be observed. We show that in contrast to polycrystalline bilayers, the loop asymmetry stays constant after multiple training cycles and that the magnitude of the asymmetry is directly correlated with the magnitude of the EB.

The exchange bias (EB) is a unidirectional magnetization anisotropy caused by exchange coupling at the interface between a ferromagnet (FM) and an antiferromagnet (AFM). Its most obvious result is a shift of the magnetic hysteresis curve after cooling through the Néel temperature of the AFM in an external magnetic field.[1-3] This shift on the magnetic field axis can be accompanied by an asymmetry of the magnetic hysteresis curve or an increased coercivity.[4-6] A gradual decrease of the EB with increasing number of field cycles is referred to as the training effect, which strongly depends on the material system.[7-9] It is attributed to irreversible changes in the AFM domain state magnetization or more generally to relaxation processes of AFM spin configurations.[10,11] In some systems, the training effects of the descending and ascending branches differ strongly, which was explained by different magnetization reversal processes in both branches of the hysteresis loop.[12-13]

Epitaxial thin film systems with cubic crystal structure usually show a twofold angular magnetic symmetry in (110) orientation, fourfold symmetry in (100) orientation, and sixfold symmetry in (111) orientation.[14] Recently, we demonstrated highly isotropic EB and coercivity in a twinned Co(111)/Co$_3$O$_4$(111) bilayer system, indicating that higher orders of lattice symmetry can lead to increasingly isotropic EB and coercivity.[15] In EB systems like Fe/MnF$_2$, different magnetization reversal processes on either side of the hysteresis loop cause a strong magnetic anisotropy[16], which is less pronounced in systems like twinned Fe/FeF$_2$(110).[17] Such magnetization reversal asymmetries are most apparent as peaks in the transverse magnetization component ($M_T$), which is accessible by magneto-optical Kerr effect (MOKE)[16,18,19] as well as anisotropic magnetoresistance (AMR)[19,20] measurements. So far, asymmetric magnetization reversal in Co/CoO bilayers has been observed for polycrystalline systems [21-23] as well as in an epitaxial thin film system with Co$_{hcp}(11\bar{2}0)$ [20] and Co$_{fcc}$(001) [21]. To the best of our knowledge, these samples have all been produced by post-oxidizing Co to CoO in-situ or in ambient conditions. A striking feature of Co/CoO systems is the reversibility of the training effect by applying a magnetic field perpendicular to the cooling field direction.[20,23] To the best of our knowledge, these peculiar effects have so far not been examined as a function of temperature and in well-defined, crystallographically characterized epitaxial systems.

Here, we investigate a Co$_{fcc}$(001)/CoO(001) bilayer system, fully epitaxially grown on MgO(001) via molecular beam epitaxy (MBE). We describe the growth of this system and provide a detailed structural and magnetic characterization. While the temperature-dependent measurements only show a slight difference between field cooling and measuring along the hard and the easy axis, all measurements below the blocking temperature reveal an asymmetric magnetization reversal. The training effect is visible on both sides of the loop and can be fitted either by the implicit Binek formula [24], a modified power law or a logarithmic function, which also works well for Co/Co$_3$O$_4$ thin films in other crystallographic orientations.[13] Angular-resolved MOKE measurements at room temperature reveal different peak widths in the transverse magnetization component, indicating varying magnetization reversal processes for different field-to-lattice orientations.

The sample was grown epitaxially by MBE on an MgO(001) substrate (dimensions 10 mm x 5 mm) in the stacking order MgO/Co/CoO. The CoO layer was grown at an oxygen partial pressure of $p$(O$_2$) = 3.3·10$^{-7}$ mbar, which is expected to result in CoO with a low defect concentration.[25] The film growth during MBE was monitored by in-situ reflection high-energy electron diffraction (RHEED) (patterns are shown in the supporting information).

The sample morphology was investigated by X-ray diffractometry (XRD) and X-ray reflectometry (XRR) using an X'Pert Pro MPD PW3040-60 diffractometer (PANalytical) with Cu $K\alpha$ radiation ($\lambda = 1.54056$ Å). The specular XRD measurement (Fig. 1a) clearly shows the Co(002) diffraction peak at $2\theta = 51.675°$, corresponding to an out-of-plane (oop) lattice constant of 3.53 Å which is close to the bulk value of 3.54 Å.[26][27] Additionally, an off-specular XRD texture map of the Co{202} peaks was measured using an Eulerian cradle. The measurement was carried out at $2\theta = 76.111°$ for 360° of sample rotation $\phi$ using a sample tilt of $\Psi = \langle 40°, 50° \rangle$. Four Co{202} peaks are visible due to the fourfold symmetry of the fcc-Co(001) lattice, indicating good crystalline growth of the Co-layer.

The crystallinity of the CoO cannot be investigated by specular XRD as its diffraction peaks overlap with the MgO substrate peaks. Therefore, a rocking-curve ($\omega$-scan) was measured around the expected position of the CoO(004) peak. As shown in Fig. 1c, a peak deconvolution of the rocking curve reveals the signal contributions of the MgO(004) substrate peak and the CoO(004) film peak at $2\theta = 93.058°$, corresponding to an oop lattice constant of 4.24 Å which is again close to previously reported values of 4.26 Å.[28] Both peaks were fitted using Lorentzian functions. An additional off-specular XRD texture map of the CoO{202} peaks at $2\theta = 61.695°$ is displayed in Fig. 1b.

The layer thickness of the stack has been determined via XRR as shown in the insert of Fig. 1c. Note that the best fit of the data was obtained when the model included a 1.6 nm interlayer of reduced density between Co and CoO, indicating a defective interface structure, likely originating from oxygen exposure of the Co layer immediately prior to the onset of CoO deposition. All three layers have a roughness below 2 nm. The full set of fit parameters and further discussion of this fit can be found in the supporting information.

High-resolution transmission electron microscopy (TEM) images confirm the epitaxial crystal structure observed by XRD (Fig. 1d), showing that the rather large lattice mismatch between Co and CoO is compensated by edge dislocations, which may contribute to a defective suboxide interface region, as indicated by XRR. The images were taken by a Jeol JEM-2200FS TEM at 200 kV from a cross-section lamella that was cut using a dual beam microscope Helios NanoLab 600i.

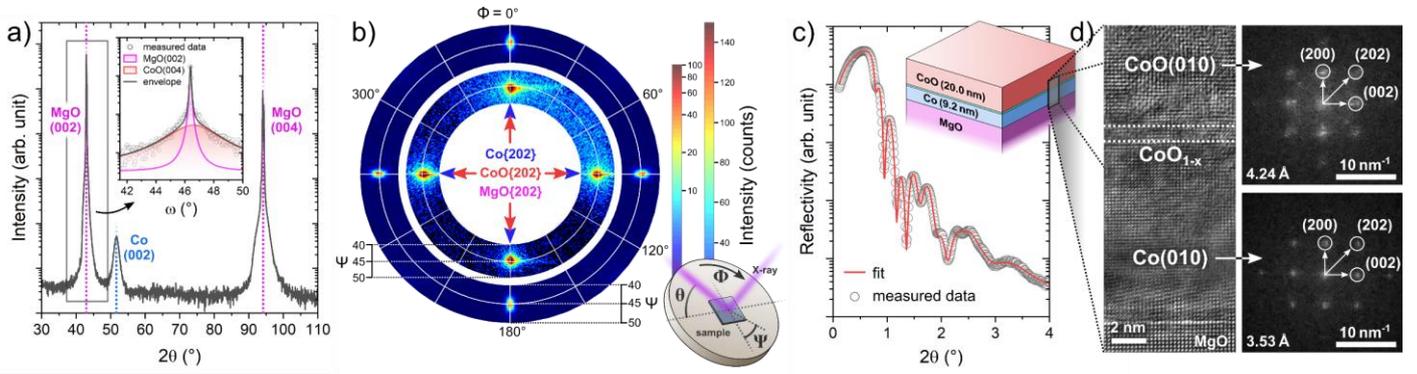

**Figure 1:** Sample characterization: **a)** Specular θ-2θ XRD scan. The inset shows a rocking curve of the CoO(004) peak. **b)** Off-specular XRD texture maps showing the Co{202} (outer ring) and CoO{202} (inner ring) peaks. **c)** XRR measurement with schematic illustration of the layer stack (a detailed discussion on the fit model can be found in the supporting information). **d)** High-resolution TEM image of the film stack, with Fast Fourier Transforms (FFT) of the CoO and Co layers, respectively. The image was then filtered using an average background subtracted high-resolution TEM (HRTEM) filter[29].

Angular-dependent room temperature investigations of the longitudinal and transverse magnetization were performed by measuring the magneto-optical Kerr effect (MOKE) using *p*-polarized light with a wavelength of 635 nm at an angle of incidence of 45° (Fig. 2). More information on the MOKE setup can be found in Ref. [30]. The longitudinal geometry is sensitive to the magnetization component parallel to the applied external magnetic field $\vec{H}$, while the transverse geometry is sensitive to the in-plane component perpendicular to $\vec{H}$. The coercivity of the longitudinal loops (Fig. 2a) only shows a slight magnetic anisotropy with a fourfold angular dependence, which can also be found in the remanence of the longitudinal and transverse magnetization component (Fig. 2b). In between the magnetic easy and hard axes, the longitudinal signal shows steps on both sides of the loops, correlated with peaks in the corresponding transverse loops (Fig. 2e). These peaks indicate in-plane rotation of the magnetization vector via stable intermediate states, i.e. neighboring easy axes that are energetically favorable for a defined field range.[16] This reversal via stable intermediate states was further investigated using vectorial MOKE magnetometry [31-33], employing measurements of linear MOKE curves with s- and p-polarized light in the longitudinal and transverse geometry. The polar plot in Fig. 2c showcases the direction and magnitude of the magnetization vector $\vec{M}$ during the magnetization reversal at a sample orientation of $\phi = -10°$ (*c.f.* Fig. 2g). Upon the decrease of $|\vec{H}|$, $\vec{M}$ orients itself along the nearest easy axis. After reversal of the applied magnetic field, $\vec{M}$ first falls into the second easy axis. The intermediate state is then characterized as the phase in which $\vec{M}$ is aligned in between the easy axis and the hard axis as indicated by the data point density in the plot. Note that the magnitude of $\vec{M}$ is reduced during the switching process, especially during the switching from one easy axis to the other, highlighting the appearance of a multidomain state in which the individual magnetic domains are not uniformly aligned. An additional video animation of the reversal process is provided in the supporting information.

The transverse magnetization component was analyzed to further characterize the switching fields, i.e., the peak positions (Fig. 2f). An inner and, for most angles, also an outer switching field (right panel of Fig. 2f) can be determined, corresponding to switching into intermediate magnetization states; the indices *a* and *d* denote the hysteresis branches for ascending and descending applied magnetic fields, respectively. Only for very broad and flat transverse loops (*e.g.*, 0° and 45° in Fig. 2e), no second switching field was observed. While the inner switching fields $H_{a1}$ and $H_{d1}$ remains approximately constant, the outer switching fields $H_{a2}$ and $H_{d2}$ show a clear fourfold in-plane symmetry, with larger values near the magnetic hard axes along Co{100} and lower values near the magnetic easy axes along Co{110}. The fourfold symmetry is also clearly visible in the peak height, which was calculated as the average height of the peak maxima in the ascending and descending branches at $H_{a1}$ and $H_{d1}$, respectively (Fig. 2g). Here, the abrupt sign switching of the peak height is located near $\phi = n \cdot 90°$, confirming the switching into and out of different easy axes, depending on the angle between $\vec{H}$ and the hard axes. Once $\vec{H}$ crosses a hard axis in $\phi$, $\vec{M}$ orients itself along the nearest neighboring easy axis during the magnetization reversal. The evaluation of both peak positions in the transversal MOKE data also gives rise to the peak widths, which corresponds to the field range over which the intermediate state is stable. The peaks get broader near the hard axes, as it is well-known from exchange bias systems such as Fe/MnF$_2$ in which these broad, nearly rectangular peaks are typical.[16,34] This is also confirmed by the fourfold symmetry of the switching fields (Fig. 2f) and the abrupt peak height switching near the hard axes (Fig. 2g).[35]

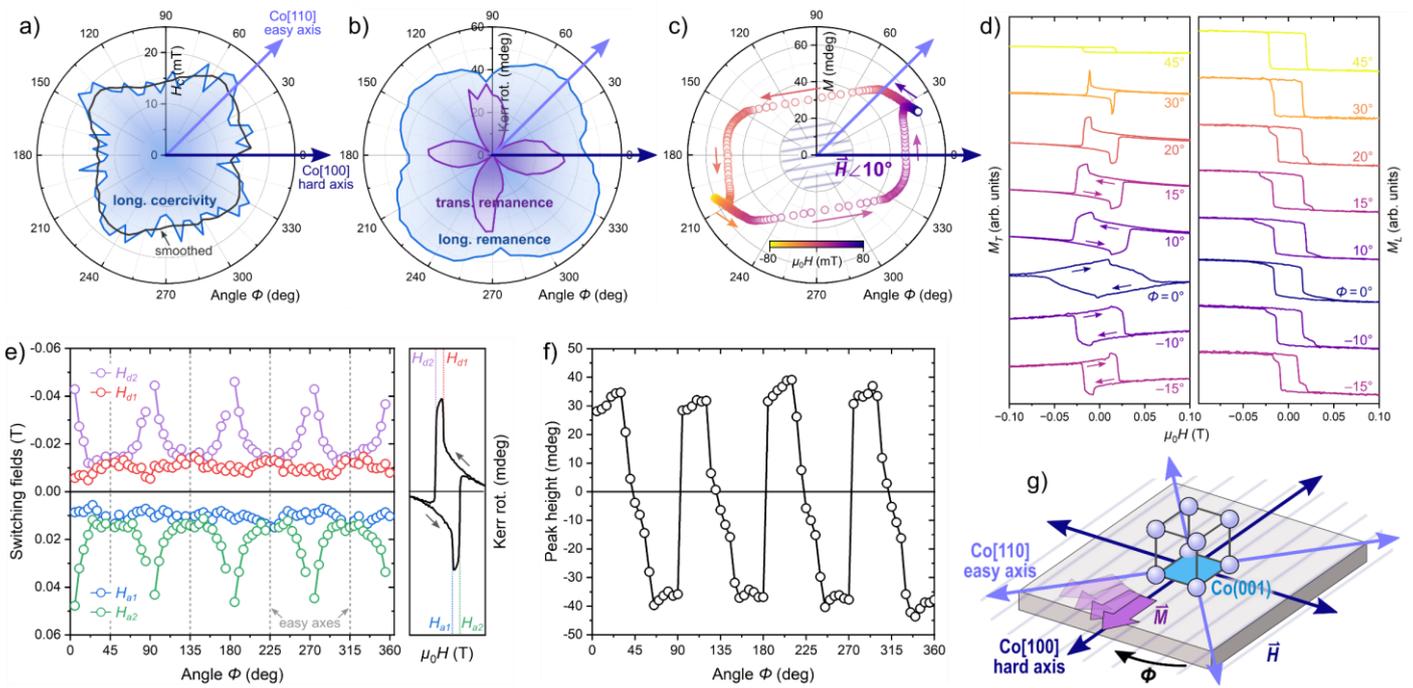

**Figure 2:** Room temperature MOKE measurements: a) Coercive fields in the longitudinal loops; b) remanence of the longitudinal and transverse magnetization component; c) vectorial MOKE magnetometry for $\phi = -10°$ showcasing the magnetization reversal including the stable intermediate state; d) exemplary transverse (left panel) and longitudinal (right panel) loops; e) inner and outer switching fields (left panel), determined from the peak positions in the transvers loops (right panel); f) peak height with respect to saturation (positive if the left peak is "up" and the right peak "down"); g) depiction of the measurement geometry.

Temperature-dependent magnetization measurements of the thin film sample were performed using the VSM mode of a MPMS 3 magnetometer (Quantum Design). The sample was cooled down from 350 K in the presence of 1 T magnetic field before starting the isothermal field loop measurements between -2 T and 2 T for each temperature. The linear diamagnetic contribution of the MgO substrate has been accounted for by subtracting a linear function. The magnetization curves were measured after field cooling along the Co[100] hard axis at $\phi = 0°$ (Fig. 3a) and along the Co[110] easy axis at $\phi = 45°$ (Fig. 3b). In both cases, the loops show a strong asymmetry below 200 K. Since these measurements are performed along the easy and hard axes, the steps seen in the longitudinal MOKE curves at intermediate angles (10°-20° off the easy axes) are barely visible.[16,17] While the loop shapes clearly differ between the two orientations, the coercive fields and the corresponding EB are of similar magnitude (Fig. 3c,d). In both cases, the EB is relatively large (about −820 Oe at $\phi = 0°$ and −760 Oe at $\phi = 45°$ at 10 K) with a blocking temperature $T_B \approx 250\ K$. This is comparable to earlier studies of epitaxial Co$_{hcp}$(11$\bar{2}$0)/CoO [20] and Co(111)/Co$_3$O$_4$(111) bilayer systems[15], indicating that the magnitude of the EB is mostly independent of crystal symmetries. Here, we quantified the loop asymmetry, as given in Fig. 3c-f, by separating two contributions of the magnetic hysteresis energy loss (*i.e.*, the area enclosed by the hysteresis loop): One is caused by normal energy dissipation $E_1$ and the other is caused by the EB $E_2$. The asymmetry is then calculated as $E_2/(E_1 + E_2) \cdot 100\%$. The method is explained in more detail in Fig. S4 in the supporting information. As displayed in Fig. 3(c,d), the loop asymmetry scales directly with the EB and therefore vanishes above $T_B$, which directly illustrates the contribution of the EB for the asymmetry of the magnetization reversal.

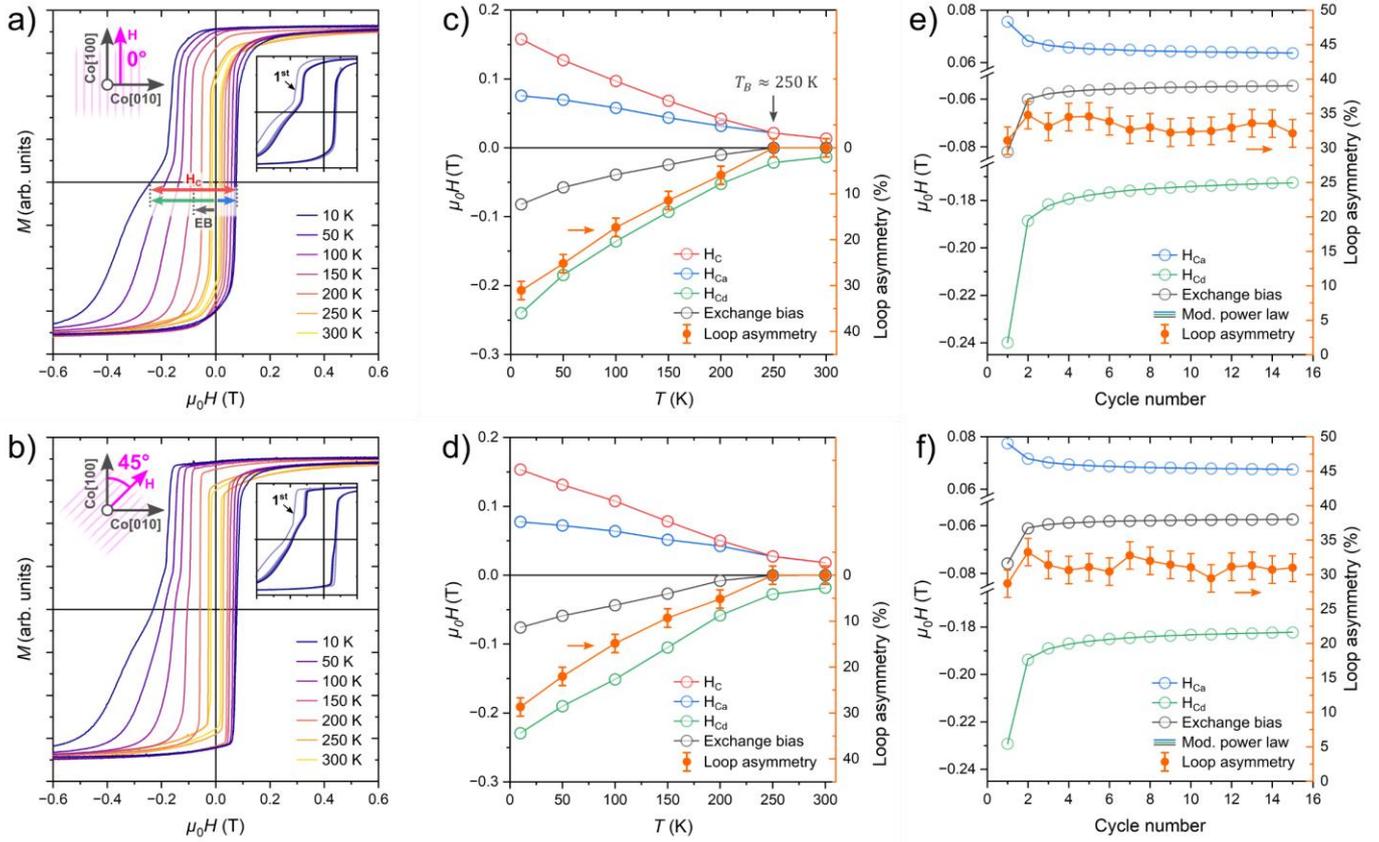

**Figure 3:** Temperature dependent VSM measurements: **a,b)** hysteresis loops for field cooling and measurement at 0° and 45° sample orientation, as indicated in the schemes; insets show the first 15 field cycles at 10 K; **c,d)** total coercive field $H_C$, coercive field of the ascending branch $H_{Ca}$, coercive field of the descending branch $H_{Cd}$, and EB, as well as loop asymmetry over temperature for field cooling and measurement at 0° (c) and 45° (d) sample orientation; the error bars indicate an estimated ±2% originating from the ambiguity in the evaluation method (see Fig. S3 in the supporting information); **e,f)** training effect and loop asymmetry as a function of field cycles at 10 K for 0° (e) and 45° (f) sample orientation.

In a previous study of epitaxial Co$_{fcc}$(001) on Si(001) with in-situ post-oxidized CoO top layer of 2-3 nm, the asymmetry has not been observed at all in magnetometry, likely because of the too large Co layer thickness.[21] Compared to the Co$_{hcp}$(11$\bar{2}$0)/CoO system [20], the loop asymmetry seen in Fig. 3a,b is much more pronounced. Additionally, it remains basically unchanged after multiple cycles (training) as shown in Figs. 3(e,f), while it has been shown to decrease in the Co$_{hcp}$(11$\bar{2}$0)/CoO system [20] and fully vanish in polycrystalline Co/CoO.[19,23] The vanishing of the asymmetry after training in polycrystalline samples has been attributed to spin reorientation in uncompensated AFM grains.[23] The training effect is then typically modeled based on the Fulcomer and Charap model, in which AFM particles/grains with uniaxial magnetic anisotropy are assumed.[36,37] In the sample presented here, the persistence of the loop asymmetry after training can be ascribed to the well-defined epitaxial CoO layer, which increases the stability of the AFM spin configuration. Based on previous calculations using a modified Fulcomer and Charap model, which accounts for magneto-crystalline anisotropy (MCA), Liu *et al.* concluded that the presence of the MCA in epitaxially grown systems leads to the survival of the loop asymmetry after training but also results in a significant reduction in the degree of asymmetry. The evaluation of the training effect of the EB, the total coercive field, as well as the coercive fields of the ascending and descending branches are shown in Fig. 3e and f. Interestingly, our epitaxial sample shows a pronounced asymmetry that is comparable to that of polycrystalline systems while fully maintaining its asymmetry after training (Fig. 3e,f), substantially deviating from the findings by Liu *et al.* Here, the data were fitted best using the modified power law, as suggested by Rui *et al.*[38]. Note that this model was developed for EB systems with spin glass (SG)/FM interfaces. Such SG/FM interfaces are known to exist also for Co/CoO systems and occur predominantly at defect-rich interfaces.[39] The CoO$_{1-x}$ interlayer between the Co and CoO, that was implied by the XRR fit and HRTEM images, therefore could be an indicator of such a SG phase. However, also the implicit Binek function and the logarithmic function provide reasonable fits. A thorough comparison of the modified power law fit with other models (a logarithmic function, a conventional power law function and the implicit Binek function[24]) is given in Fig. S3 in the supporting information.

In summary, we observed a magnetization reversal via stable intermediate states as well as an asymmetric magnetization reversal below the blocking temperature in exchange biased epitaxial fcc-Co(001)/CoO(001). The sample was grown on MgO(001) via MBE. The exchange bias showed a blocking temperature near 250 K and was measured along the Co[100] hard axis and the Co[110] easy axis. A large loop asymmetry was observed for both orientations. Due to the well-defined, epitaxial CoO layer, the loop asymmetry remained stable after training. The temperature-dependent measurement of the magnetization curves revealed that the strength of the loop asymmetry is directly proportional to the magnitude of the EB. At room temperature, above $T_B$, MOKE measurements show a magnetization reversal via stable intermediate states which could be used to build quaternary memory devices or other spintronics devices with more than two stable states.

## Acknowledgement

We gratefully acknowledge the valuable contributions of Tapas Samanta, who was involved in earlier stages of this work but could not be reached for co-authorship at the time of submission. We thank Prof. Günter Reiss from Bielefeld University for making available laboratory equipment.

## Competing interests

The authors declare no conflict of interest.

## Author contributions

Conceptualization: A.E., M.G., and M.W.; Investigation: A.E., F.P., H.T., I.E., J.B., M.G., M.W., and T.S.; Formal analysis: A.E., F.P., J.B., J.F., M.G., and M.W.; Software: A.E. and J.F.; Visualization: M.W.; Resources, Supervision, Funding acquisition, and Writing – review & editing: A.E., A.H., L.C., M.W., and T.B.; Writing – original draft: A.E., J.B., M.G., and M.W.

## Data availability statement

The data that support the findings of this study are available from the corresponding author upon reasonable request.